\documentclass[twocolumn,epjc3]{svjour3}
\smartqed  
\RequirePackage{graphicx}

\journalname{Eur. Phys. J. C}
\usepackage{color}
\usepackage{graphicx}
\usepackage{dcolumn}
\usepackage{bm}
\usepackage{color}
\usepackage{enumitem}
\usepackage{amsmath}
\usepackage{amssymb}
\usepackage{orcidlink}
\usepackage{graphicx} 
\usepackage{cuted} 
\usepackage{doi}
\usepackage{cite}
\usepackage{hyperref}
\usepackage{soul}
\hypersetup{
  colorlinks=true,        
  linkcolor=blue,         
  citecolor=cyan,         
}

\begin{document}

\title{Can a Dehnen-type dark matter halo affect the neutrino flavor oscillations?}

\author{Mirzabek Alloqulov \thanksref{e1,addr2, addr3,addr1}  \orcidlink{0000-0001-5337-7117}  \and Ahmadjon Abdujabbarov \thanksref{e2, addr1,addr4,addr6} \orcidlink{0000-0002-6686-3787} \and Bobomurat~Ahmedov~\thanksref{e3, addr1,addr5,addr6} \orcidlink{0000-0002-1232-610X} \and Chengxun Yuan \thanksref{e4, addr1} \orcidlink{0000-0002-2308-6703}}
\thankstext{e1}{malloqulov@gmail.com (corresponding author)}
\thankstext{e2}{ahmadjon@astrin.uz}
\thankstext{e3}{ahmedov@astrin.uz}
\thankstext{e4}{yuancx@hit.edu.cn (corresponding author)}

\institute{University of Tashkent for Applied Sciences, Str. Gavhar 1, Tashkent 100149, Uzbekistan \label{addr2} \and New Uzbekistan University, Movarounnahr str. 1, Tashkent 100000, Uzbekistan \label{addr3}  \and School of Physics, Harbin Institute of Technology, Harbin 150001, People’s Republic of China \label{addr1} \and Tashkent State Technical University, Tashkent 100095, Uzbekistan \label{addr4} \and Institute for Advanced Studies, New Uzbekistan University, Movarounnahr str. 1, Tashkent 100000, Uzbekistan \label{addr5} \and Institute of Theoretical Physics, National University of Uzbekistan, Tashkent 100174, Uzbekistan \label{addr6} }

\date{Received: date / Accepted: date}

\maketitle

\begin{abstract}
The gravitational weak lensing of neutrinos in the presence of a Schwarzschild black hole surrounded by a Dehnen-type dark matter halo is investigated. The event horizon structure is explored, and the existence of the BH is studied in two different slices of the parameter space. Additionally, we derive analytic expressions for the oscillation phase and transition probabilities for both radial and non-radial neutrino propagation. Finally, we use a two-flavor toy model, and numerically analyze how the dark matter halo parameters as density and scale radius, affect oscillation probabilities and examine the role of decoherence. The results show that the presence of a dark matter halo modifies the oscillation phase and damping factor, leading to measurable deviations from the standard Schwarzschild BH case. These findings suggest that neutrino oscillations could, in principle, serve as a probe for dark matter distributions around compact astrophysical objects.
\\

\end{abstract}

\maketitle

\section{Introduction}
In general relativity (GR), black holes (BHs) arise as exact analytical solutions of Einstein’s field equations. The first and most relevant  solutions with spherical and axial symmetry correspond to Schwarzschild~\cite{1916SPAW.......189S,2015arXiv151202061B} and Kerr BHs~\cite{1963PhRvL..11..237K}, respectively. Since then, numerous additional solutions have been developed within various extensions and modifications of gravity theories. As a result, BHs are commonly classified according to their mass scales: stellar-mass BHs ($3$--$100\,M_\odot$), intermediate-mass BHs ($10^2$--$10^5\,M_\odot$), and supermassive BHs (SMBHs) ($10^6$--$10^{10}\,M_\odot$). Beyond these, there are also hypothetical mini BHs, often referred to as primordial BHs, which are expected to have formed from density fluctuations during the early stages of the Universe.  

Over the years, GR has been rigorously tested in both weak- and strong-field regimes. These include, for instance, the direct imaging of SMBH shadows such as M87$^*$~\cite{Akiyama19L1,Akiyama19L6} and SgrA$^*$~\cite{Event}, together with the groundbreaking detection of gravitational waves~\cite{Abbott_2016}. Yet, despite such achievements, our direct knowledge of the Universe accounts for only about $4\%$ of its total content. Observations of the cosmic microwave background reveal that $27\%$ consists of dark matter (DM) and $69\%$ of dark energy, thereby motivating detailed studies of BHs embedded in DM halos. The distribution and structure of DM halos play an essential role in shaping galactic rotation curves and in astrophysical events such as the Bullet Cluster collision. Although DM is generally assumed to interact solely via gravity, compelling observational evidence continues to confirm its presence. Consequently, several analytic halo models that incorporate BHs have been introduced, including the Einasto~\cite{Merritt_2006,Dutton_2014}, Navarro–Frenk–White~\cite{Navarro_1996}, Burkert~\cite{Burkert_1995} and Dehnen profiles~\cite{Dehnen93,refId0,Gohain_2024,Pantig_2022}. More recently, the effects of Dehnen-type DM halos on BH geometries have been explored from different perspectives~\cite{errehymy2025,2025arXiv250911245S,2025arXiv250908569A,2025NuPhB101817069B,2025EPJC...85..798A,2025EPJC...85..677R,2025arXiv250515540L,2025ChPhC..49e5101A,2025CoTPh..77c5402A,2024PDU....4601683G}. In particular, a modified Schwarzschild solution surrounded by a Dehnen-type DM halo has recently been derived in~\cite{Al-Badawi_2025}.

We aim to investigate the effect of the DM halo on the oscillations of neutrinos. Neutrino oscillations describe the process in which neutrinos, elementary particles that interact only weakly with matter, change their flavor as they propagate through spacetime. In the presence of a gravitational field, i.e. within curved spacetime, this phenomenon is influenced by the background geometry. In such situations, the equations governing neutrino motion explicitly depend on the curvature, resulting in modifications to their trajectories and energies. These changes can, in principle, manifest themselves as observable variations in oscillation behavior. Consequently, the transition probabilities between different neutrino flavors, namely electron, muon, and tau, differ from those in flat spacetime. This possibility has motivated during the last years numerous theoretical studies exploring neutrino oscillations in curved geometries~\cite{Cardall:1996cd,Piriz:1996mu,Ahluwalia:1996ev,Bhattacharya:1999na,Pereira:2000kq,Crocker:2003cw,Lambiase:2005gt,Godunov:2009ce,Ren:2010yf,Geralico:2012zt,Chakraborty:2013ywa,Visinelli:2014xsa,Zhang:2016deq,Alexandre:2018crg,Blasone:2019jtj,Buoninfante:2019der,Boshkayev:2020igc,Mandal:2021dxk,Koutsoumbas:2019fkn,Wudka:1991tg,Fornengo_1997,Swami_2020,Swami_2021,Swami:2022xet,Chakrabarty:2021bpr,Chakrabarty:2023kld,Alloqulov_2025,Alexandre:2025qip,Shi_2025,Shi2025,wang2025}.

The influence of spacetime geometry on neutrino propagation and oscillation necessitates a careful examination of the modifications introduced by dark matter halos. Analyzing neutrino oscillations in curved spacetime provides a crucial framework for understanding their behavior in extreme astrophysical contexts, such as during supernova explosions or in the intense gravitational fields of compact, massive objects. Such investigations are expected to yield valuable insights into both the fundamental characteristics of neutrinos and the manner in which they interact with gravitational phenomena \cite{Swami_2020,Chakrabarty:2021bpr}. Our investigation commences with a brief review of the spacetime of the Schwarzschild BH surrounded by a Dehnen-type DM halo. Furthermore, we consider neutrino oscillations in flat spacetime. We obtain the analytic expression for the phase of the neutrino oscillations in curved spacetime by considering the radial and non-radial propagation of neutrinos. Following this, we analyze the effect of spacetime parameters on neutrino oscillation probability. We also investigate the phenomenon of decoherence.  

The article is structured as follows: In Section~\ref{sec:2}, we briefly review the spacetime of the Schwarzschild BH surrounded by a Dehnen-type DM halo, including the region of the BH and No-BH and the structure of the event horizon. Section~\ref{sec:3} represents oscillations of neutrinos in flat spacetime. The phase of neutrino oscillations in curved spacetime is obtained by considering two scenarios of neutrino propagation in Section~\ref{sec:4}. In Section~\ref{sec:5}, we calculate the probability of neutrino oscillation and explore the effect of the DM halo considering a two flavor toy model. Section~\ref{sec:6} is devoted to the decoherence properties. Our conclusions and a discussion of future prospects are presented in Section~\ref{sec:summary}.

\section{Spacetime}\label{sec:2}

\begin{figure*}
\centering
\includegraphics[scale=0.5]{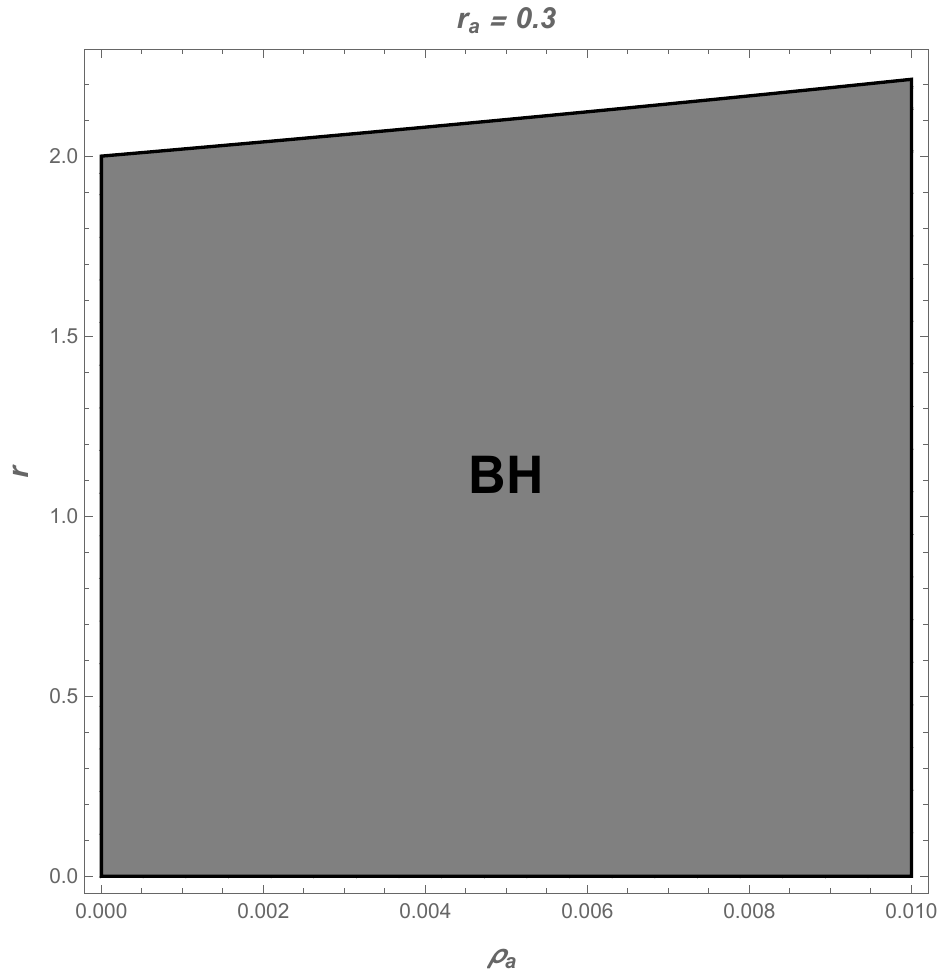}
\includegraphics[scale=0.5]{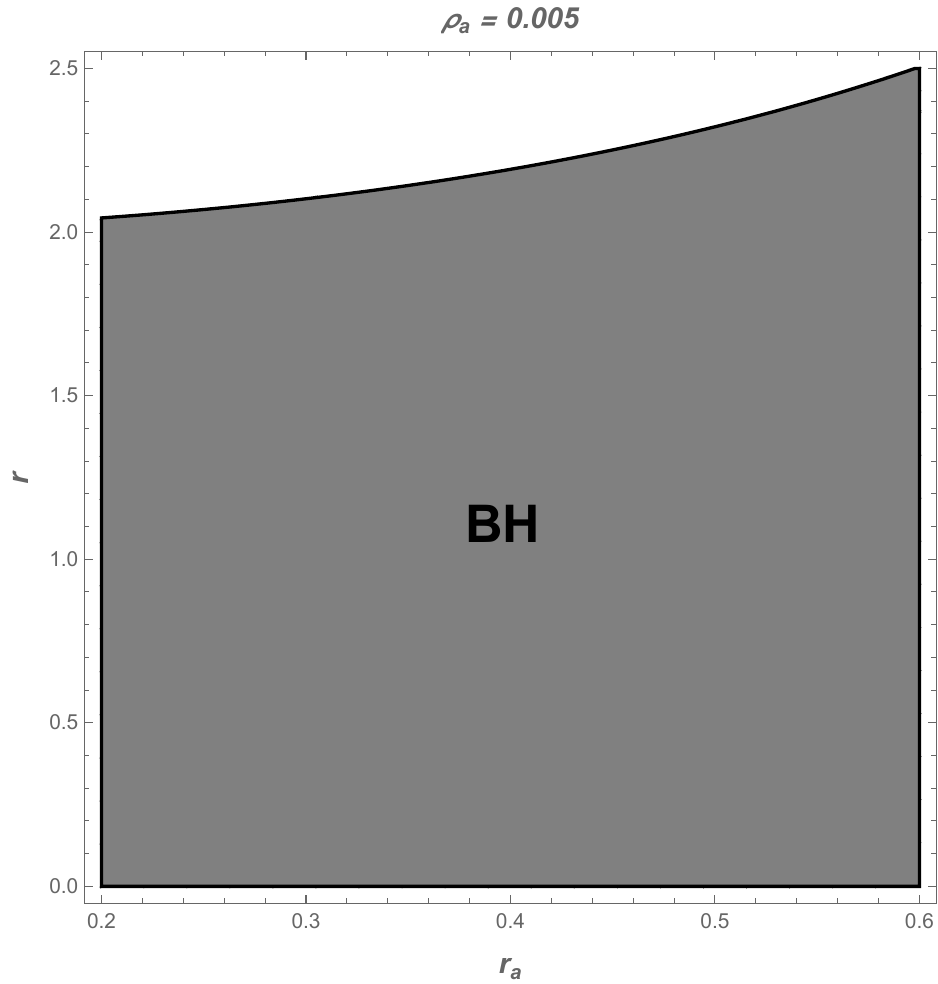}
\caption{Left panel: The phase diagram demonstrates the existence of the Schwarzschild BH surrounded by a Dehnen-type DM halo region in the $(\rho_a,r)$ plane. The black region represents the slice of the parameter space where $f(r,\rho_a) \leq 0$, pointing to the presence of the Schwarzschild BH surrounded by a Dehnen-type DM halo. Right panel: The phase diagram illustrates the existence of the Schwarzschild BH surrounded by a Dehnen-type DM halo region in the $(r_a,r)$ plane. The black region represents the slice of the parameter space where $f(r,r_a) \leq 0$, pointing to the presence of the Schwarzschild BH surrounded by a Dehnen-type DM halo. The boundary lines separate the BH and No-BH regions.}
\label{fig:region}
\end{figure*}

\begin{figure}
    \centering
    \includegraphics[scale=0.5]{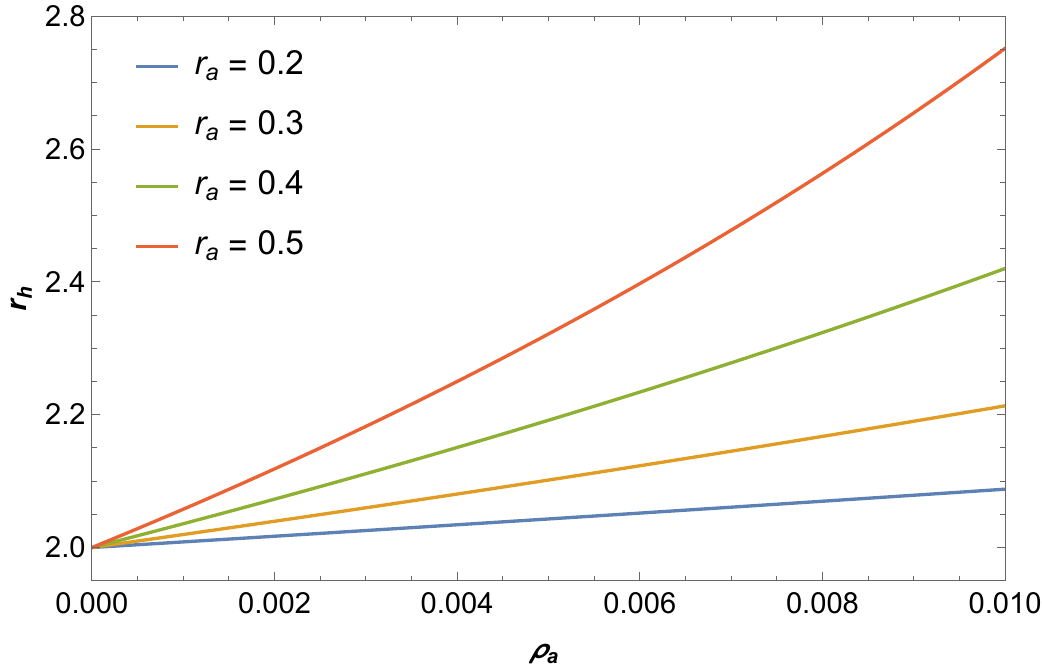}
    \caption{The plot shows the event horizon radius of the Schwarzschild BH surrounded by a Dehnen-type DM halo as a function of $\rho_a$ for the different values of the $r_a$. }
    \label{fig:horizon}
\end{figure}

In this section, we briefly review the spacetime of the Schwarzschild BH surrounded by a Dehnen-type DM halo. Recently, the solution was obtained in Ref.~\cite{Al-Badawi_2025}. According to this, we can write the line element of the Schwarzschild BH surrounded by a Dehnen-type DM halo as~\cite{Al-Badawi_2025}
\begin{equation}
    ds^2 = -f(r)dt^2 + \frac{dr^2}{f(r)} + r^2(d\theta^2 + \sin^2\theta d\phi^2),
\end{equation}
with
\begin{equation}
f(r)=1-\frac{2M}{r}-32\pi \rho_a r^3_a\sqrt{\frac{r+r_a}{r^2_a r}}\,,
\end{equation}
where $r_a$ and $\rho_a$ refer to the radius and density of the DM halo. It should be noted that Schwarzschild BH can be recovered when $\rho_a=0$. Our investigation begins with exploring the existence of BH for the different parameters. Fig.~\ref{fig:region} shows the phase diagram in two different slices of the parameter space, which are $(\rho_a,r)$ and $(r_a,r)$, showing the regions where a Dehnen-type DM halo surrounds the Schwarzschild BH. The black areas represent the slice of the parameter space where $f(r,\rho_a)\leq 0$ and $f(r,r_a)\leq 0$. Boundary lines separate the BH and No-BH areas. In addition, the effect of the parameters of the DM halo on the event horizon radius is shown in Fig.~\ref{fig:horizon}. It can be observed from this figure that the values of the event horizon radius increase with an increase in the parameters of the DM halo. 

We can rename the line element as follows
\begin{equation}
ds^2=-{\cal A}dt^2+{\cal B}dr^2+{\cal C}d\theta^2+{\cal D}d\phi^2
\end{equation}
where
\begin{align}
g_{tt}=-{\cal A}\,, \quad g_{rr}={\cal B}\,, \quad g_{\theta\theta}={\cal C}\,, \quad g_{\phi\phi}={\cal D}
\end{align}
In the next sections, we consider the above line element for simplicity. 

\section{Neutrino oscillations in flat spacetime}\label{sec:3}

In this section, we investigate the neutrino oscillations in flat spacetime. Note that neutrinos are produced and detected by different flavor eigenstates in weak interactions, which is $|\nu_{\alpha} \rangle$ where $\alpha=e,\mu, \tau$. In addition, the flavor eigenstates can be written as a superposition of the mass eigenstates as follows
\begin{equation}
|\nu_{\alpha}\rangle=\sum_i U_{\alpha i}^{*}|\nu_{i}\rangle\,,
\end{equation}
where $i$ runs from 1 to 3, and $U$ refers to the unitary mixing matrix of $3\times3$. If we consider a three flavor neutrino oscillation, this matrix represents the Pontecorvo–Maki–Nakagawa–Sakata leptonic mixing matrix~\cite{Pontecorvo:1957qd,10.1143/PTP.28.870,Pontecorvo:1967fh}. Here, we make the assumption that the neutrino wave function is a plane wave (see Refs.~\cite{Pontecorvo:1957qd,10.1143/PTP.28.870,Pontecorvo:1967fh}), and it is emitted from a source $S$ at $(t_S,x_S)$ and is later detected at $(t_D,x_D)$. After that, one can write the wave function at the detector point as
\begin{equation}
|\nu_i(t_D,x_D)\rangle=\exp{(-i\Phi_i)|\nu_i(t_S,x_S)\rangle}\,,
\end{equation}
where $\Phi_i$ refers to the oscillation phase. If we consider neutrinos to be produced in $|\nu_{\alpha}\rangle$ at $S$ and then detected in $|\nu_{\beta}\rangle$ at $D$, we can write the probability of the change as
\begin{eqnarray}
    {\cal P}_{\alpha \beta}&=&|\langle\nu_{\beta}|\nu_{\alpha}(t_D,x_D)\rangle |^2 =\sum_{i,j}U_{\beta i}U^{*}_{\beta j}U_{\alpha j}U^{*}_{\alpha i}\times \nonumber \\
    && \times \exp{(-i(\Phi_i-\Phi_j))}.
\end{eqnarray}
It is worth noting that a flavor change is possible provided that $\Phi_i \neq \Phi_j$. Since neutrino mass eigenstates have different masses and energy/momentum, they accumulate different phases ($\Phi_i$) during propagation. As a result, it leads to an increase in neutrino oscillation phenomena~\cite{Akhmedov2009}. The phase in flat spacetime can be written as follows
\begin{equation}\label{phase}
\Phi_i=E_i(t_D-t_S)-{\bf p}_i \cdot ({\bf x}_D-{\bf x}_S).
\end{equation}
A typical assumption is that all mass eigenstates within a flavor eigenstate are initially produced at $S$ with the same momentum or energy~\cite{Akhmedov2009,Akhmedov2011}, and $(t_D-t_S)\simeq|{\bf x}_D-{\bf x}_S|$ for relativistic neutrinos. After that, one can write the phase difference as
\begin{equation}
    \Delta\Phi_{ij}=\Phi_i-\Phi_j\simeq\frac{\Delta m^2_{ij}}{2E_0}|x_D-x_S|\,,
\end{equation}
where $E_0$ represents the average energy of the relativistic neutrinos produced at the source, and $\Delta m^2_{ij}=m^2_i-m^2_j$. Moreover, we can rewrite Eq.~(\ref{phase}) in covariant form for the propagation of neutrinos in curved spacetime as
\begin{equation}
    \Phi_i= \int_S^D p^{(i)}_{\mu} d x^{\mu},
\end{equation}
with
\begin{equation}\label{p(i)}
    p^{(i)}_{\mu}=m_i g_{\mu \nu} \frac{d x^{\nu}}{d s},
\end{equation}
represents the conjugate canonical momentum to the coordinate $x^{\nu}$ and $ds$ and $g_{\mu \nu}$ refer to the line element and metric tensor, respectively.

\begin{figure*}[!htb]
    \centering
    \includegraphics[scale=0.5]{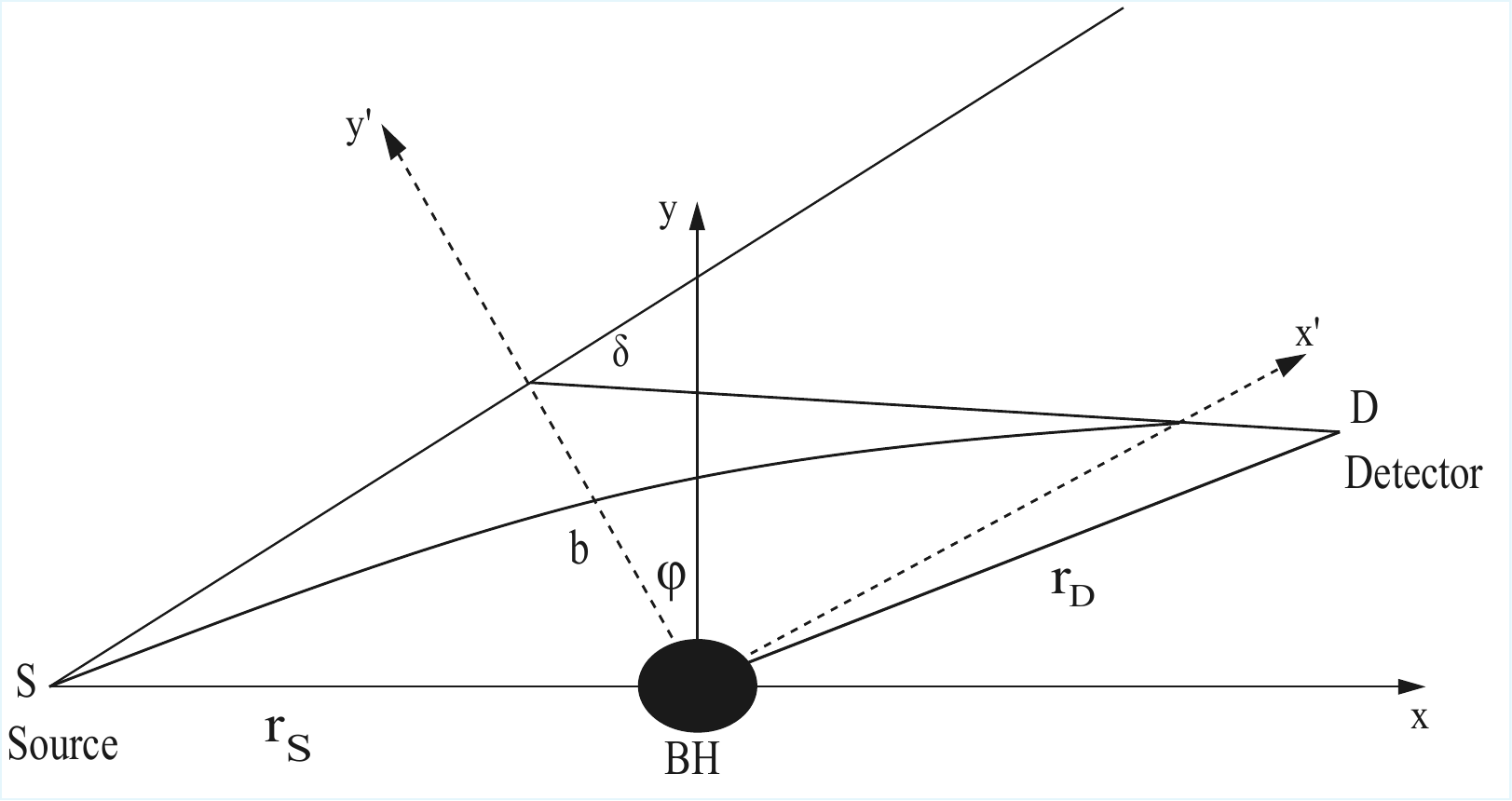}
    \caption{The plot demonstrates the schematic diagram of the gravitational weak lensing of neutrinos in the spacetime of the Schwarzschild BH surrounded by a Dehnen-type DM halo.}
    \label{fig:diagram}
\end{figure*}

\section{Phase of lensed neutrinos}\label{sec:4}

In this section, we calculate the phase of the neutrino oscillation considering two types of neutrino propagation. Firstly, we need to find the components of the canonical momenta $p_{\mu}^{(k)}$ for the test particles. Note that $k$ is responsible for the neutrino mass eigenstates. We consider the motion of the test particles on the equatorial plane, i.e. $\theta=\pi/2$. $p^{(k)}_t$ and $p^{(k)}_{\phi}$ are constant along the trajectory of particles because the components of the metric tensor do not depend on $t$ and $\phi$. The constants of motion can be written as $p^{(k)}_t=-E_k$ and $p^{(k)}_{\phi}=J_k$, and it is possible to write $p^{(k)}_{r}$ as $p_k$ for simplicity.  Using the mass-shell relation, we have the following expression
\begin{equation}\label{Eq:mass-shell}
    -m_k^2=g^{tt} E_k^2+g^{rr} p_k^2+g^{\phi \phi} J_k^2.
\end{equation}
We can calculate the phase of neutrino oscillations in the following subsections.

\subsection{Radial propagation}

In this part, we explore the radial propagation of neutrinos in the equatorial plane. Therefore, $J_k=0$ and $\dot{\theta}=0$. We can write the following relations using Eq.~(\ref{p(i)})
\begin{align}\label{relations}
\dfrac{d t }{d s}=-\dfrac{E_k}{m_k g_{tt}}, \quad \dfrac{d r }{d s}=\dfrac{p_k}{m_k g_{rr}}.
\end{align}
Using the above equations, we can write the phase of the neutrino oscillations as follows
\begin{equation}
    \Phi_k= \int_S^D \left[-E_k \left(\dfrac{d t}{d r}\right)_0+p_k(r)\right]dr,
\end{equation}
where the subscript $0$ denotes the light-ray trajectory. We can obtain the light-ray differential from Eq.~(\ref{relations}) as
\begin{equation}\label{differential}
    \left(\dfrac{dt}{dr}\right)_0=\dfrac{E_0}{p_0(r)}\dfrac{\mathcal{B}}{\mathcal{A}},
\end{equation}
From Eq.~(\ref{Eq:mass-shell}), one can get the following equation
\begin{eqnarray}\label{momenta}
   && p_k(r)=\pm \sqrt{\dfrac{\mathcal{B} E_k^2}{\mathcal{A}}-\mathcal{B} m_k^2}\,.
\end{eqnarray}
Then the expression for the phase yields as
\begin{equation}
     \Phi_k= \pm \int_S^D E_k \sqrt{\dfrac{\mathcal{B}}{\mathcal{A}}}\left[-1+\sqrt{1-\dfrac{m_k^2 \mathcal{A}}{E_k^2}}\right]dr.
\end{equation}
The expression for the phase can be rewritten by expanding the square root under the bracket as
\begin{equation}
     \Phi_k= \pm \int_S^D \sqrt{\mathcal{A}\mathcal{B}} E_k \dfrac{m_k^2}{2 E_k^2} dr\,.
\end{equation}
If we consider the relativistic approximation ($m_k<<E_k$)~\cite{Fornengo_1997}, one can get the following
\begin{eqnarray}
  && E_k \simeq E_0 + \mathcal{O}\left(\dfrac{m_k^2}{2 E_0}\right), \nonumber \\
  && E_k \dfrac{m_k^2}{2 E_k^2} \simeq E_0 \dfrac{m_k^2}{2 E_0^2},
\end{eqnarray}
and the neutrino oscillation phase becomes as
\begin{equation}
  \Phi_k= \pm \dfrac{m_k^2}{2 E_0}\int_S^D  \sqrt{\mathcal{A} \mathcal{B}} d r.
\end{equation}
We consider the Schwarzschild BH surrounded by a Dehnen-type DM halo. Therefore, ${\cal A B}=1$, and we get the same result for the phase as in the Schwarzschild BH case. The phase of radial propagation of neutrinos~\cite{Fornengo_1997,Swami_2020} is
\begin{equation}\label{eq:r1}
    \Phi_k \approx \pm \dfrac{m_k^2}{2 E_0}\left|r_D-r_S\right|\,.
\end{equation}

\subsection{Non-radial propagation}

Here, we investigate the non-radial propagation of neutrinos on the equatorial plane, i.e., $\dot{\theta}=0$ and $J_k \ne 0$. The phase for this case is
\begin{equation}
    \Phi_k=\int_S^D \left[-E_k\left(\dfrac{d t}{d r}\right)_0 + p_k + J_k \left(\dfrac{d \phi}{d r}\right)_0\right] d r,
\end{equation}
where $J_k$ refers to the angular momentum of the $k$-th mass eigenstate of the neutrino, and one can define $\Big(\frac{dt}{dr}\Big)_0$ and $\Big(\frac{d\phi}{dr}\Big)_0$ along the light-ray trajectories as
\begin{align}\label{Eq:light-ray}
    \left(\dfrac{dt}{dr}\right)_0=\dfrac{E_0}{p_0}\dfrac{\mathcal{B}}{\mathcal{A}} , \quad \left(\dfrac{d\phi}{dr}\right)_0=\dfrac{J_0}{p_0}\dfrac{\mathcal{B}}{\mathcal{D}}.
\end{align}
$J_k$ can be expressed as a function of energy $E_k$ in the following form
\begin{equation}\label{Eq:angular}
    J_k=E_k b v_k^{\infty},
\end{equation}
where $v_k^{\infty}$ and $b$ refer to the velocity at infinity and the impact parameter, respectively. The following relations can be written due to the asymptotically flat spacetime 
\begin{eqnarray}\label{Eq:vinfty}
    && v_k^{\infty}=\dfrac{\sqrt{E_k^2-m_k^2}}{E_k} \simeq 1-\dfrac{m_k^2}{2 E_k^2}, \nonumber \\
    && J_k \simeq E_k b \left(1-\dfrac{m_k^2}{2 E_k^2}\right),
\end{eqnarray}
Note that the relativistic approximation is considered in the above relations. The angular momentum of a massless particle is $J_0=E_0b$. After using the above equations, we can get the expression for the phase as
\begin{equation}
    \Phi_k = \int _S^D \dfrac{E_0 E_k \mathcal{B}}{p_0}\left[-\dfrac{1}{\mathcal{A}}+\dfrac{p_0 p_k}{E_0 E_k \mathcal{B}}+\dfrac{b^2}{\mathcal{D}}\left(1-\dfrac{m_k^2}{2 E_k^2}\right)\right]d r.
\end{equation}
We can simplify the above integral by using the mass-shell relation. Firstly, we write the following expression through Eq.~(\ref{Eq:mass-shell}) as follows
\begin{equation}\label{Eq:relations}
    \dfrac{p_0(r) p_k(r)}{E_0 E_k \mathcal{B}}=\dfrac{1}{\mathcal{A}}-\dfrac{b^2}{\mathcal{D}}-\dfrac{m_k^2}{2 E_k^2},
\end{equation}
The above equation is valid for the massless case $k=0$ and $m_0=0$. Finally, we obtain the phase expression by using the relativistic approximation as
\begin{eqnarray}
        \Phi_k &=& -\dfrac{m_k^2}{2 E_0} \int_S^D \dfrac{E_0 \mathcal{B}}{p_0}d r = \nonumber \\ 
        &=& \pm \dfrac{m_k^2}{2 E_0} \int_S^D \sqrt{\mathcal{A} \mathcal{B}}\left(1-\dfrac{b^2 \mathcal{A}}{\mathcal{D}}\right)^{-1/2}d r.
    \label{Eq:phase2}
\end{eqnarray}
The above integral describes the phase of the non-radially propagating neutrinos in the spacetime of the Schwarzschild BH surrounded by a Dehnen-type DM halo. In fact, we can analyze the above integral in two scenarios. Our first case involves a neutrino non-radially propagating outward after escaping the Schwarzschild BH surrounded by a Dehnen-type DM halo metric's potential. By using the weak-field approximation, we can get the following equation

\begin{eqnarray}
&&\Phi_k=\pm \frac{m_k^2}{2E_0}\int_S^D \Big[\frac{1}{\sqrt{\frac{-b^2+r^2 + 32 b^2 r_a^2 \pi \rho _a}{r^2}}}- \nonumber \\
&&-\frac{b^2 \left(8 \pi  r_a^3 \rho_a+1\right)M/r}{\left(-b^2+r^2 + 32 \pi  b^2 r_a^2 \rho_a\right) \sqrt{\frac{-b^2+r^2 +32 \pi  b^2 r_a^2 \rho_a}{r^2}}}\Big]dr
\end{eqnarray}
The above integral can be integrated analytically. It is

\begin{strip}
 \begin{eqnarray}
\Phi_k&=&\pm\frac{m_k^2}{2E_0}\Big[\sqrt{r_D^2-b^2 \left(1-32 \pi  \rho _a r_a^2\right)}-\sqrt{r_S^2-b^2 \left(1-32 \pi  \rho _a r_a^2\right)}+\frac{M \left(1+8 \pi  \rho _a r_a^3\right)}{1-32 \pi  \rho _a r_a^2} \times \nonumber \\
&\times&\Big(\frac{r_D}{\sqrt{r_D^2-b^2 \left(1-32 \pi  \rho _a r_a^2\right)}}-\frac{r_S}{\sqrt{r_S^2-b^2 \left(1-32 \pi  \rho _a r_a^2\right)}}\Big)\Big]
\end{eqnarray}   
\end{strip}
If we consider $\rho_s=0$, we can recover the result for Schwarzschild BH~\cite{Fornengo_1997}.  

In the second case, neutrinos are emitted by a distant source and pass in the vicinity of the source of the Schwarzschild BH surrounded by a Dehnen-type DM halo. We can assume that a neutrino travels from its source $S$, past a gravitational lens described by the  Schwarzschild BH surrounded by a Dehnen-type DM halo at its closest approach point $C$, and finally to a detector $D$ as demonstrated in Fig.~\ref{fig:diagram}. Therefore, the phase integral in Eq.~\eqref{Eq:phase2} can be divided into two parts: the path from the source $S$ to the point of closest approach $C$, and the path from $C$ to the detector $D$ as follows
\begin{eqnarray}
&& \Phi_k=\dfrac{m_k^2}{2 E_0}\Bigg[ \int_{r_C}^{r_S} \sqrt{\dfrac{\mathcal{A} \mathcal{B}}{\left(1-\dfrac{b^2 \mathcal{A}}{\mathcal{D}}\right)}} dr + \nonumber \\
&& +\int_{r_C}^{r_D} \sqrt{\dfrac{\mathcal{A} \mathcal{B}}{\left(1-\dfrac{b^2 \mathcal{A}}{\mathcal{D}}\right)}} dr\Bigg],
\label{Eq:phase3}
\end{eqnarray}
where $r_S$, $r_D$ and $r_C$ refer to the distances from the lens corresponding to the source, detector and closest approach point, respectively. The distance of the closest approach point can be calculated as
\begin{equation}\label{Eq:closest1}
    \left(\dfrac{d r}{d \phi}\right)_0=\dfrac{p_0(r_C) \mathcal{D}}{J_0 \mathcal{B}}=0,
\end{equation}
We can find it by using the weak-field approximation in the following form
\begin{equation}
r_C \approx b\Big(\sqrt{1-32\pi\rho_ar_a^2}-\frac{M+8\pi\rho_ar_a^2}{b}\Big)\,.
\end{equation}
Note that we can recover the result for the Schwarzschild BH when $\rho_a=0$. 

To calculate the phase for the neutrino oscillations, we firstly expand the integrand as follows~\cite{Alexandre:2025qip}
\begin{eqnarray}\label{integrand}
  \sqrt{\frac{\mathcal{AB}}{1-{b^2 \mathcal{A}}/{\mathcal{D}}}} &\simeq& \frac{1}{\sqrt{1-{b^2}/{r^2}(1-32\pi \rho_a r^2_a)}}-\nonumber \\
 &&  - \frac{b^2M (1+8\pi \rho_a r^3_a)}{[r^2-b^2(1-32\pi \rho_a r^2_a)]^{3/2}}+\,\,...\,.  
\end{eqnarray} 
After that, we calculate the integral separately as
\begin{eqnarray}
&&\int^r_{r_C}\frac{du}{\sqrt{1-{b^2}/{u^2}(1-32\pi \rho_a r^2_a)}}=\nonumber \\
&& \qquad \ \sqrt{r^2-b^2(1-32\pi \rho_a r^2_a)}- \nonumber \\
&& \qquad -\sqrt{r_C^2-b^2(1-32\pi \rho_a r^2_a)}\,.
\end{eqnarray}
It can be easily seen by considering the distance of the closest approach point that the expression in the second square root is negative. Therefore, the result contains real and imaginary parts as follows
\begin{eqnarray}
&&\int^r_{r_C}\frac{du}{\sqrt{1-{b^2}/{u^2}(1-32\pi \rho_a r^2_a)}}=\nonumber \\
&& \qquad \sqrt{r^2-b^2(1-32\pi \rho_a r^2_a)}\pm \nonumber \\
&& \qquad \pm i\sqrt{|r_C^2-b^2(1-32\pi \rho_a r^2_a)|}\,.
\end{eqnarray}
We can calculate the integral of the second term of Eq.~(\ref{integrand}) by performing the same approach as
\begin{eqnarray}
&&-b^2 M (1+8\pi \rho_a r^3_a)\int^r_{r_C}\frac{du}{[u^2-b^2(1-32\pi \rho_a r^2_a)]^{3/2}}=\nonumber\\
&&=\frac{M(1+8\pi\rho_ar_a^3)}{1-32\pi \rho_a r^2_a}\times\Bigg[\frac{r}{\sqrt{r^2-b^2(1-32\pi \rho_a r^2_a)}}\pm \nonumber\\
&&\pm \frac{ir_C}{\sqrt{|r_C^2-b^2(1-32\pi \rho_a r^2_a)|}}\Bigg]
\end{eqnarray}
Note that the imaginary parts cancel out between the approaching and receding regimes (see,~\cite{Alexandre:2025qip}), and we can get the final expression for the phase as

\begin{eqnarray}\label{Eq:phase4}
\Phi_k&\approx&\frac{m_k^2}{2E_0}\Big[\sqrt{r_D^2-b^2(1-32\pi\rho_ar_a^2)}+ \nonumber \\
&+&\sqrt{r_S^2-b^2(1-32\pi\rho_ar_a^2)}+\frac{M \left(1+8 \pi  \rho _a r_a^3\right)}{1-32 \pi  \rho _a r_a^2} \times \nonumber \\
&\times&\Big(\frac{r_D}{\sqrt{r_D^2-b^2 \left(1-32 \pi  \rho _a r_a^2\right)}}+ \nonumber \\
&+& \frac{r_S}{\sqrt{r_S^2-b^2 \left(1-32 \pi  \rho _a r_a^2\right)}}\Big)\Big]\ .
\end{eqnarray}
The above expression can be simplified by assuming $b<<r_{S,D}$. Then it takes the following form
\begin{eqnarray}
\Phi_k&\approx&\frac{m_k^2}{2E_0}(r_S+r_D)\Big[1-\frac{b^2}{2 r_Sr_D}(1-32 \pi  \rho _a r_a^2)+ \nonumber \\
&+&\frac{2M}{r_S+r_D}\frac{\left(1+8 \pi  \rho_a r_a^3\right)}{1-32 \pi  \rho_a r_a^2}\Big]\,.
\end{eqnarray}
Again, we can recover the result for the Schwarschild BH setting $\rho_a=0$~\cite{Fornengo_1997}.

\begin{figure*}[!htb]
\centering
\includegraphics[scale=0.45]{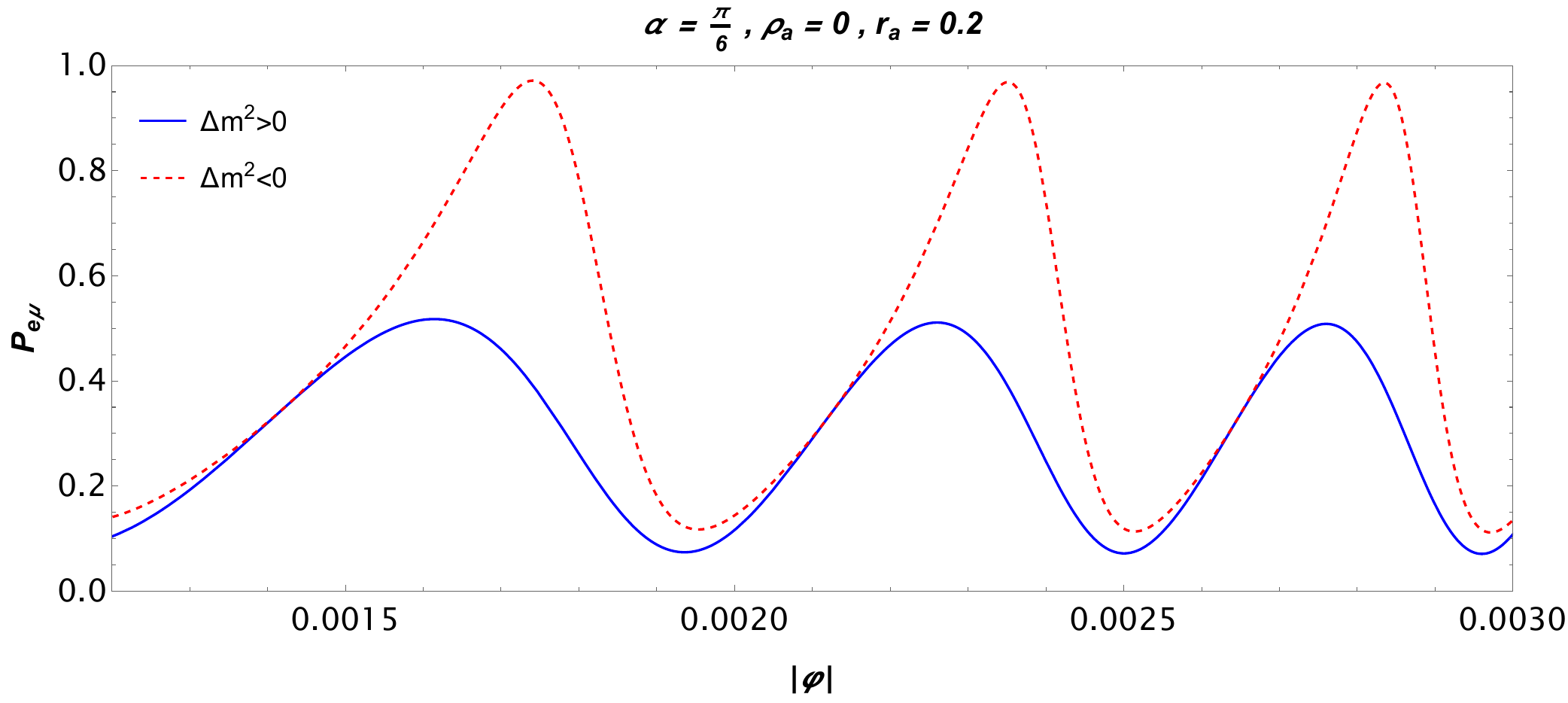}
\includegraphics[scale=0.45]{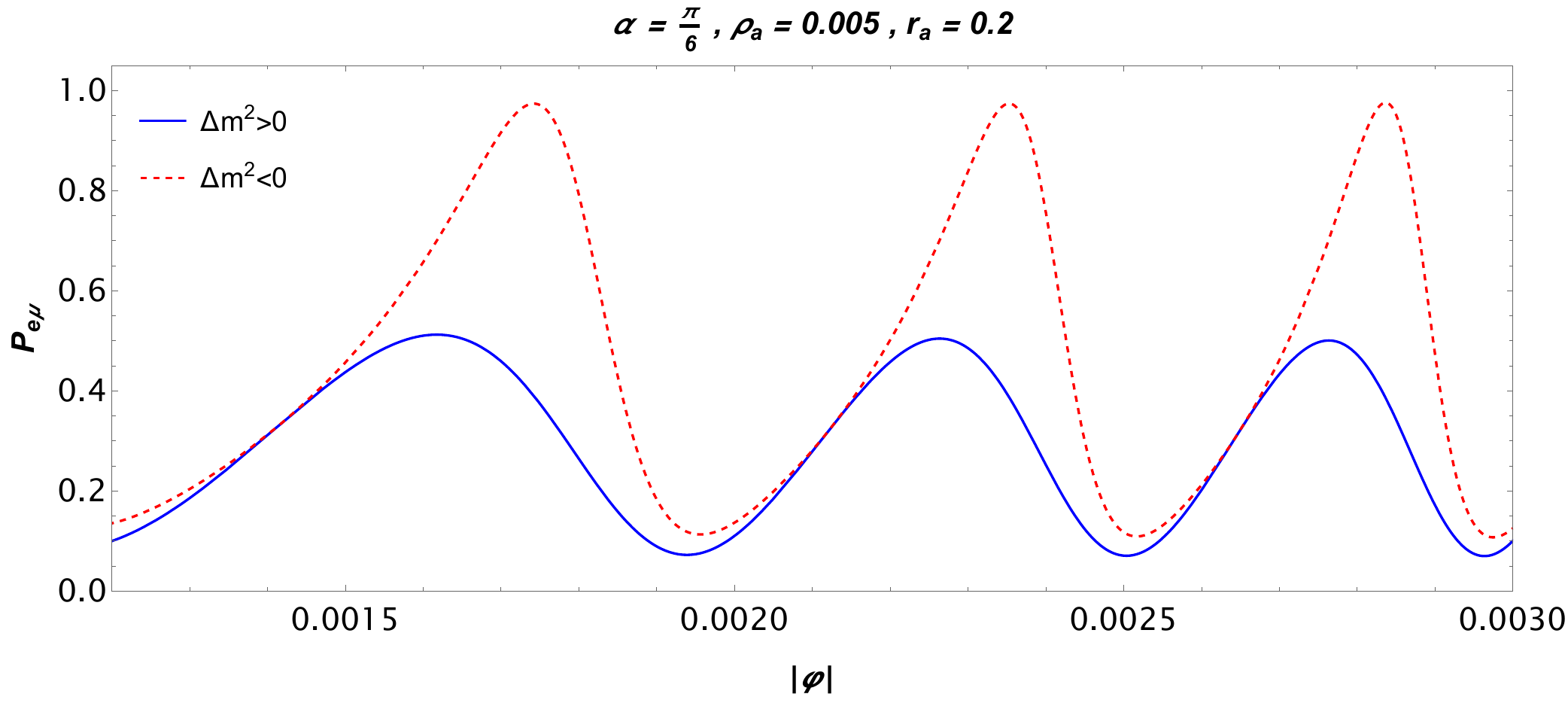}
\includegraphics[scale=0.45]{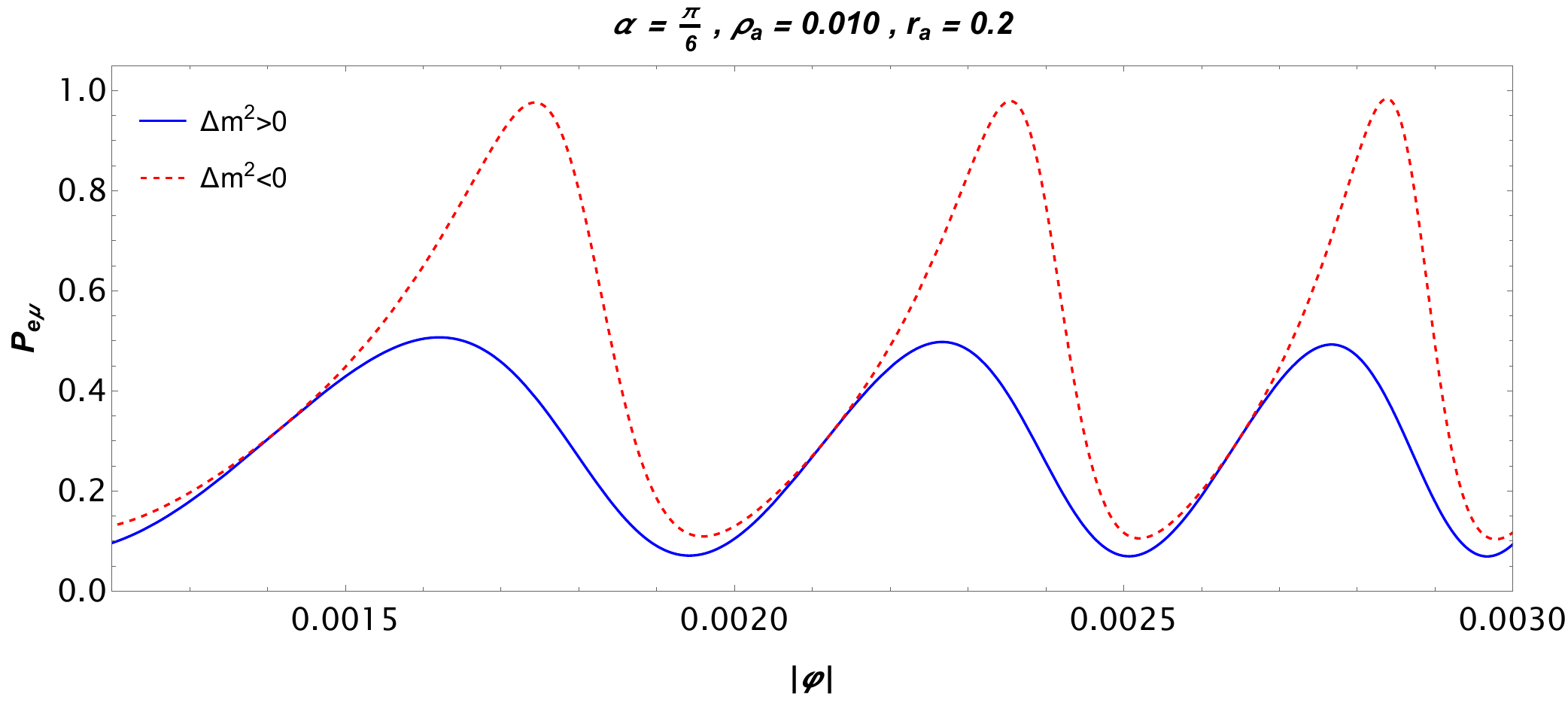}
\caption{The plot demonstrates the probability of neutrino oscillation as a function of azimuthal angle $\varphi$. From top to bottom, the densities of the central halo are $\rho_a=0$, $\rho_a=0.005$, and $\rho_a=0.01$, respectively. Blue and red-dashed lines correspond to the normal hierarchy $\Delta m^2 >0$ and inverted hierarchy $\Delta m^2 <0$, respectively. The other parameters are as follows: the mixing angle $\alpha=\pi/6$, $r_a=0.2$, $M=1M_{\odot}$, $\Delta m^2=10^{-3} eV^2$. Here, we consider that the lightest neutrino is massless. }
\label{fig:prob1}
\end{figure*}
\begin{figure*}[!htb]
\centering
\includegraphics[scale=0.45]{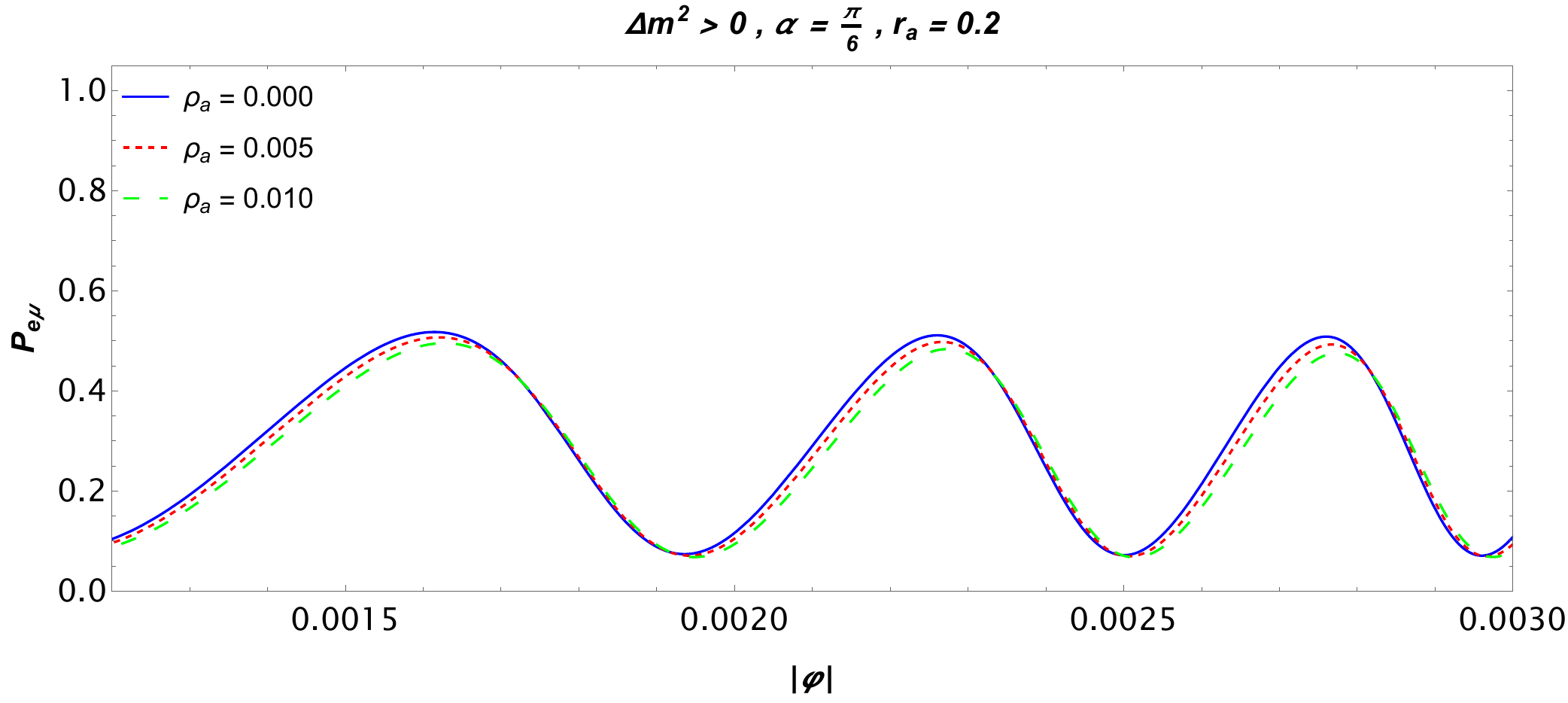}
\includegraphics[scale=0.45]{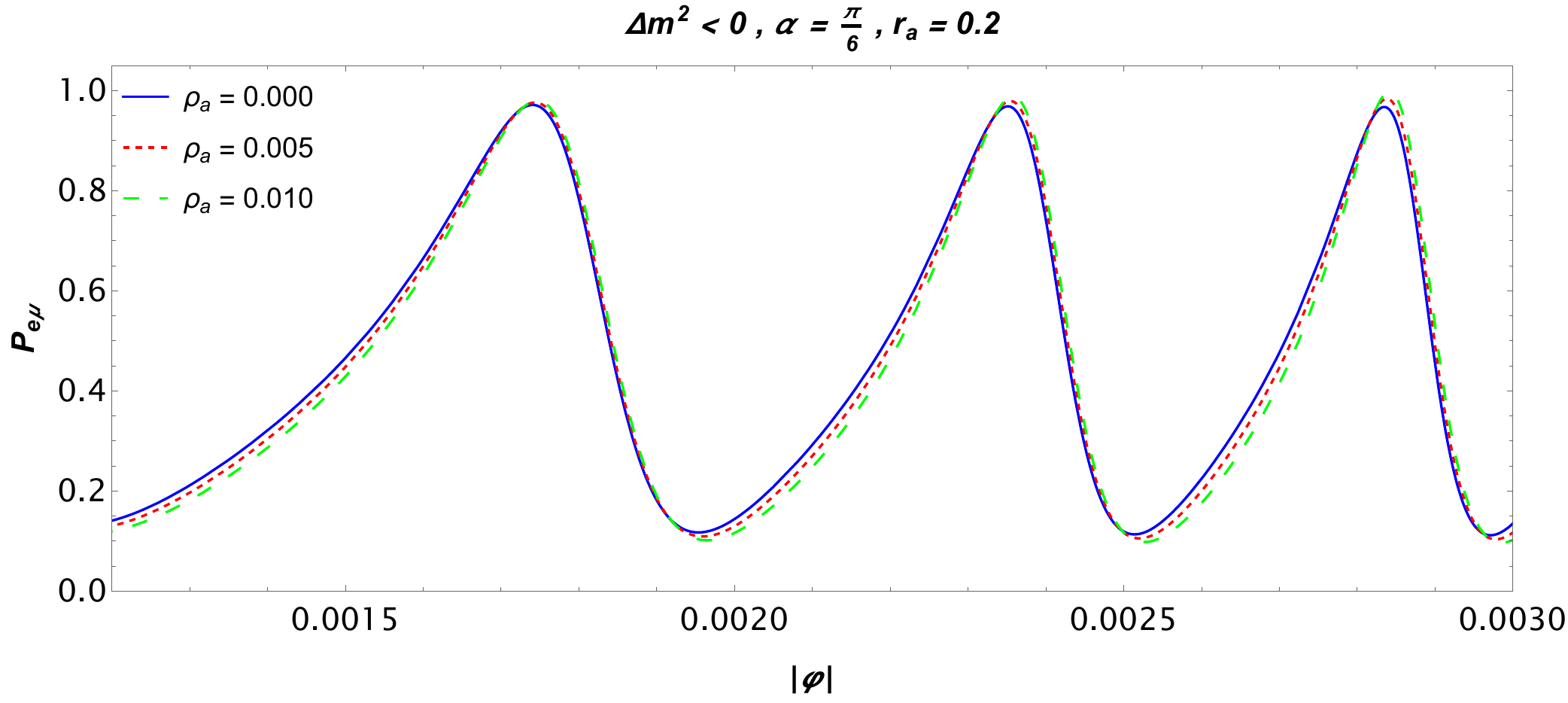}
\caption{The plot shows the neutrino oscillation probability as a function of azimuthal angle $\varphi$ for different values of the density of the DM halo. Top and bottom panels correspond to normal hierarchy $\Delta m^2>0$  and inverted hierarchy $\Delta m^2<0$, respectively. The other parameters are as follows: the mixing angle $\alpha=\pi/6$, $r_a=0.2$, $M=1M_{\odot}$, $\Delta m^2=10^{-3} eV^2$. Here, we consider that the lightest neutrino is massless. }
\label{fig:prob2}
\end{figure*}
\begin{figure*}[!htb]
\centering
\includegraphics[scale=0.45]{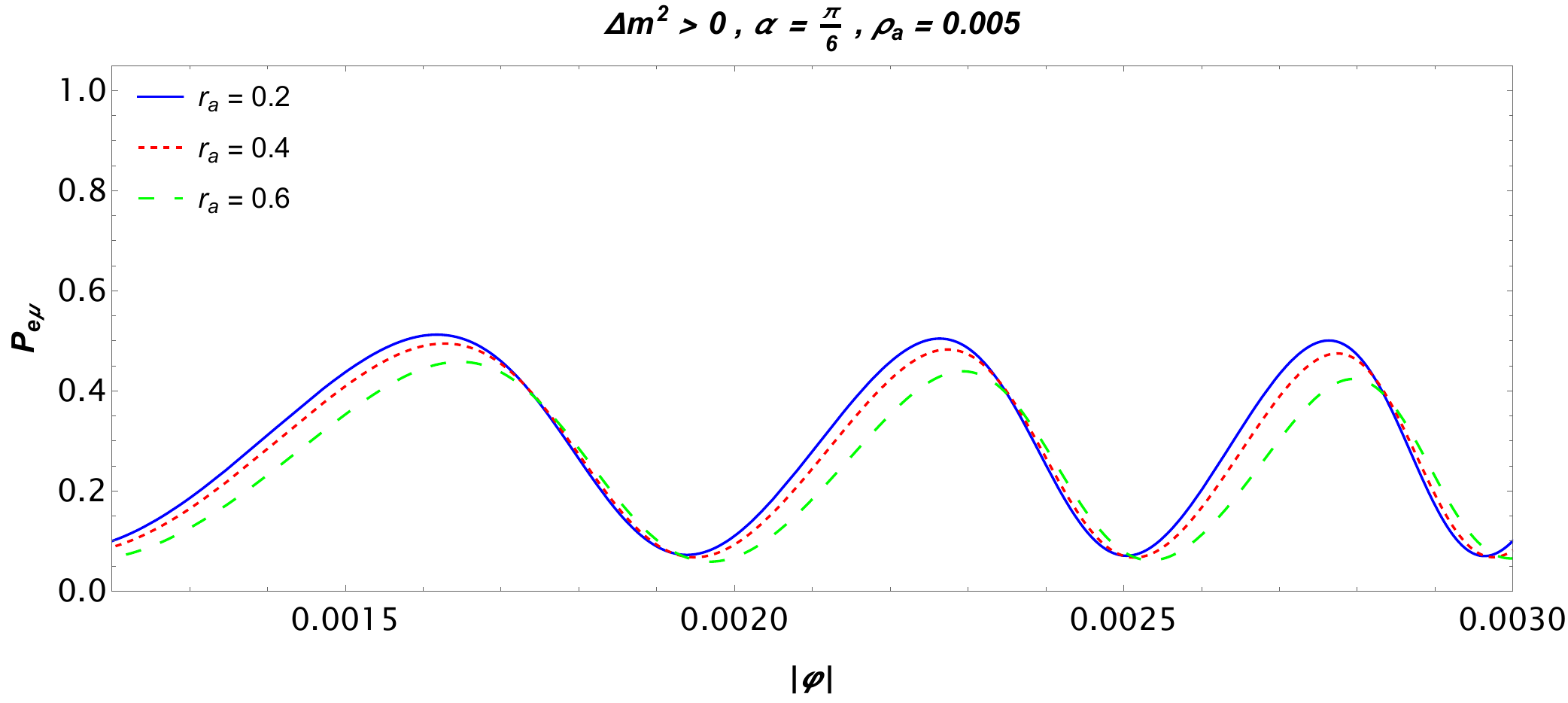}
\includegraphics[scale=0.45]{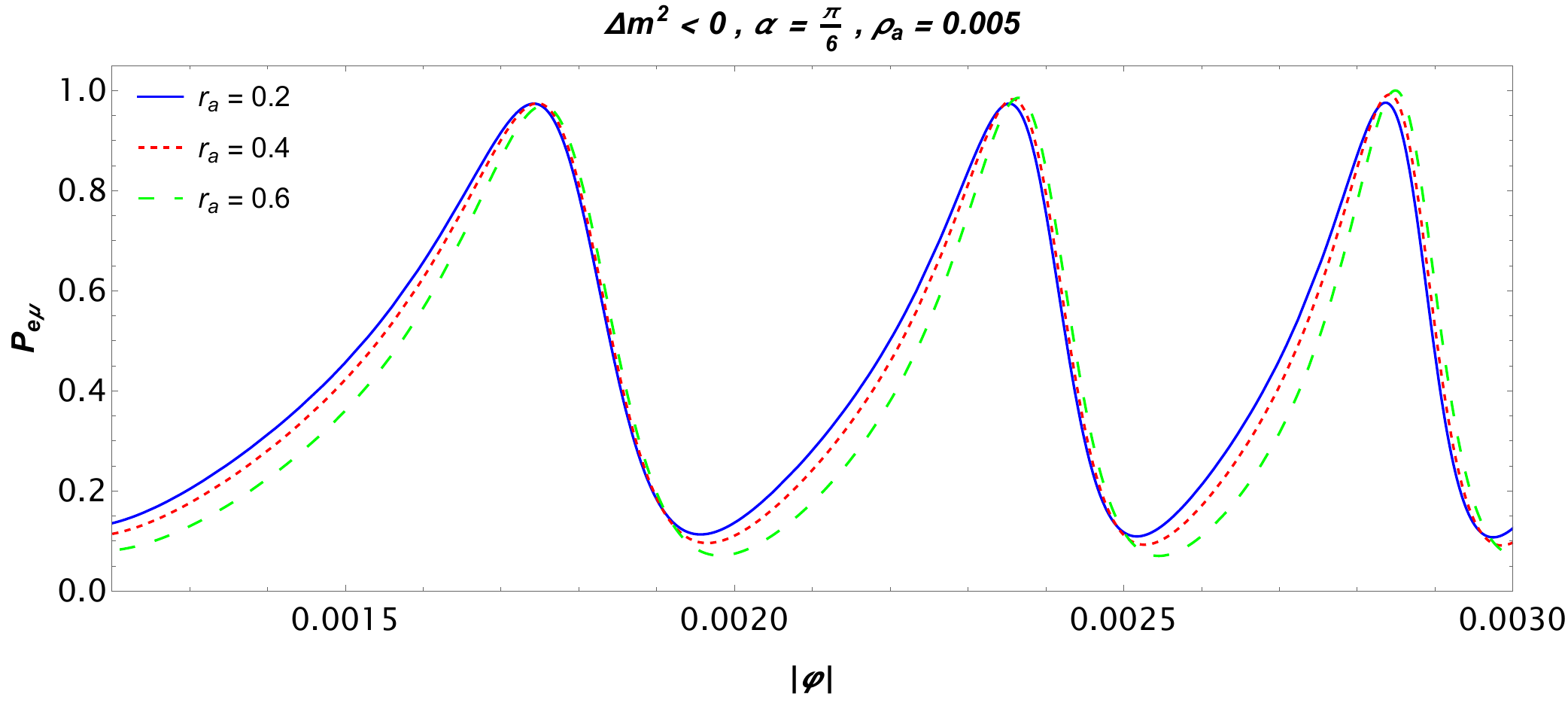}
\caption{The plot shows the neutrino oscillation probability as a function of azimuthal angle $\varphi$ for different values of the radius of the DM halo. Top and bottom panels correspond to normal hierarchy $\Delta m^2>0$  and inverted hierarchy $\Delta m^2<0$, respectively. The other parameters are as follows: the mixing angle $\alpha=\pi/6$, $\rho_a=0.005$, $M=1M_{\odot}$, $\Delta m^2=10^{-3} eV^2$. Here, we consider that the lightest neutrino is massless.}
\label{fig:prob3}
\end{figure*}
\begin{figure*}[!htb]
\includegraphics[scale=0.45]{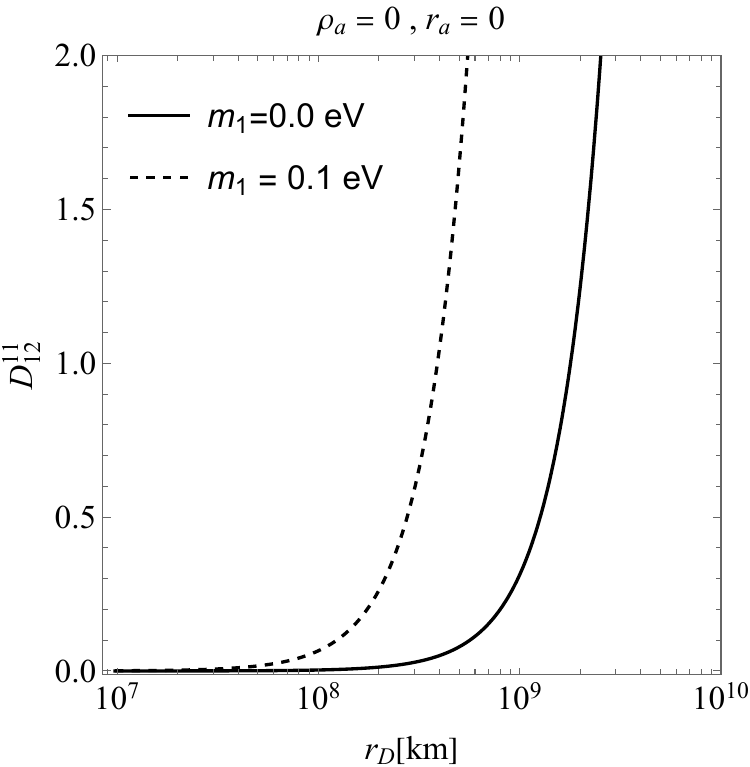}
\includegraphics[scale=0.45]{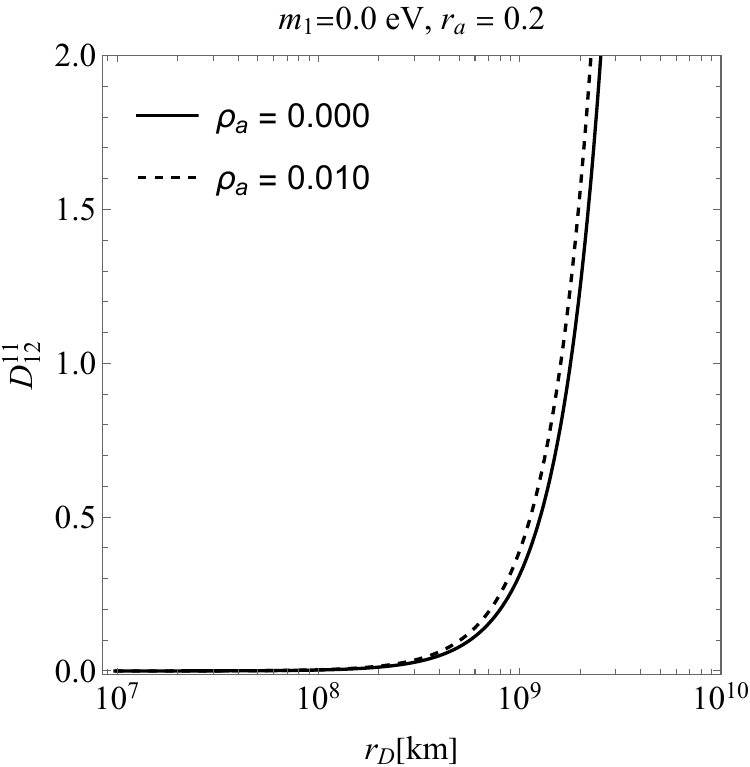}
\includegraphics[scale=0.45]{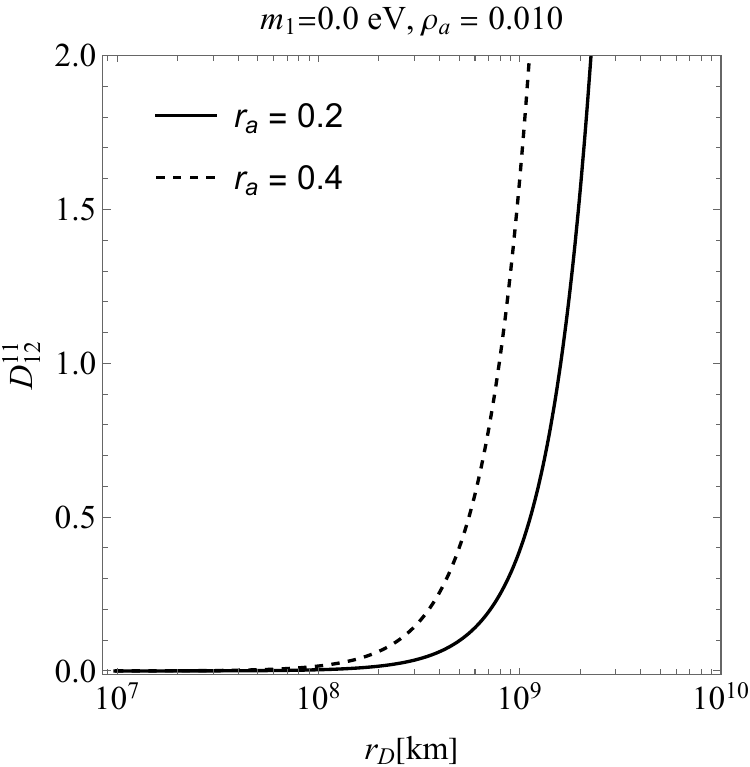}
\caption{The plot demonstrates the dependence of the damping factor $D_{12}^{11}$ on the $r_D$ for $\Bar{\sigma}/E_{loc}=10^{-13}$. Left panel: The black solid and dashed lines correspond to the $m_1=0\ eV$ and $m_1=0.1\ eV$, respectively. The middle and right panels were plotted for different values of the density and radius of the DM halo. Here, we set the other parameters for all panels as: $r_S=10^5r_D$, $\Delta m_{21}^2=m_2^2-m_1^2=10^{-3}\ eV^2$ and $E_{loc}=10MeV$. }
\label{fig:damping}
\end{figure*}

\section{Probabilities of neutrino oscillation}\label{sec:5}

In this section, we can calculate the oscillation probabilities for the gravitationally lensed neutrino using the phase we have derived. Neutrinos with mass eigenstates $\nu_i$ travelling in the spacetime of the Schwarzschild BH surrounded by a Dehnen-type DM halo are considered. A neutrino can travel from source $S$ to detector $D$ along two paths, $p$ and $q$, the difference in the proper distance leading to quantum interference. At the source, a neutrino is produced in a flavor eigenstate as 
\begin{equation}
    |\nu_{\alpha}, S \rangle = \cos\theta \, |\nu_1\rangle + \sin\theta \, |\nu_2\rangle,
\end{equation}
propagates to the detector as 
\begin{equation}
    |\nu_{\alpha}, D \rangle = N \sum_i U^{*}_{\alpha i} \sum_p \exp\!\left(-i \Phi^p_i \right) |\nu_i\rangle,
\end{equation}
where $N$ denotes the normalization factor and $\Phi^p_i$ corresponds to the phase specified in Eq.~\eqref{Eq:phase4}. Here, the impact parameter $b_p$ should be regarded as path-dependent, i.e., determined by the trajectory $p$. The probability of a neutrino flavor change from $\nu_{\alpha}$ to $\nu_{\beta}$ at the detector can be written in the following form
\begin{equation}\label{Eq:probability1}
 \mathcal{P}_{\alpha\beta}=|\langle\nu_{\beta}|\nu_\alpha,D \rangle|^2 =  |N|^2 \sum_{i,j}  U_{\beta i} U^{*}_{\beta j} U_{\alpha j}U^{*}_{\alpha i}\sum_{p,q}\exp{(-i \Delta \Phi^{pq}_{ij})},
\end{equation}
with
\begin{equation}
    |N|^2=\left(\sum_i |U_{\alpha i}|^2 \sum_{p,q} \exp(-i \Delta \Phi^{pq}_{ii})\right)^{-1},
\end{equation}
refers to the normalization factor, and $\Delta \Phi_{ij}^{pq}$ represents the phase difference as
\begin{equation}\label{Eq:phasedif}
    \Phi^{pq}_{ij}=\Phi_i^p-\Phi_j^q=\Delta m_{ij}^2 A_{pq}+\Delta b_{pq}^2 B_{i j},
\end{equation}
where
\begin{eqnarray}\label{Eq:AB}
   A_{pq}&=&\dfrac{r_S+r_D}{2 E_0}\Big[1+\frac{2M}{r_S+r_D}\frac{\left(1+8 \pi  \rho_a r_a^3\right)}{1-32 \pi  \rho_a r_a^2}- \nonumber\\
   &-&\frac{\Sigma b_{pq}^2}{4 r_Sr_D}(1-32 \pi  \rho _a r_a^2)\Big], \nonumber \\
  B_{ij}&=&-\dfrac{\Sigma m^2_{ij}}{8 E_0}\left(\dfrac{1}{r_S}+\dfrac{1}{r_D}\right)\,,
\end{eqnarray}
where $\Delta m_{ij}^2$, $\Sigma m_{ij}^2$, $\Delta b_{pq}^2$ and $\Sigma b_{pq}^2$ can be defined as
\begin{eqnarray}
    && \Delta m^2_{ij}=m^2_{i}-m^2_{j},\hspace{0.5 cm} \Sigma m^2_{ij}=m^2_{i}+m^2_{j}, \nonumber\\
    &&\Delta b^2_{pq}=b^2_{p}-b^2_{q},\hspace{0.75cm} \Sigma b^2_{pq}=b^2_{p}+b^2_{q}.
\end{eqnarray}
It should be noted that $A_{pq}$ and $B_{ij}$ do not change under the interchange of their indices, i.e., $A_{pq}=A_{qp}$ and $B_{ij}=B_{ji}$. In contrast, the oscillation probability is influenced by the total sum of the neutrino mass squares, $\Sigma m^2_{ij}$, through the parameter $\Delta b^2_{pq}$. For trajectories where $\Delta b^2_{pq}$ vanishes, the probability $\mathcal{P}^{\text{lens}}_{\alpha \beta}$ remains invariant under the transformation of the change $m_i^2 \rightarrow m_i^2 + C$. In contrast, for nonzero trajectories $\Delta b^2_{pq}$, this symmetry is broken. Under such a shift, the coefficients transform as $B_{ij} \rightarrow B_{ij} + 2C$. One can get the following expression by combining the above equations
\begin{eqnarray}
     {\mathcal P^{lens}_{\alpha\beta}}&=&|N|^2 \Big[\sum_{i,j}U_{\beta i}U^{*}_{\beta j}U_{\alpha j}U^{*}_{\alpha i}\Big(\sum_{p=q}\exp{(-i\Delta m^2_{ij}A_{pp})}+\nonumber \\
    &+& 2 \sum_{p>q} \cos{(\Delta b^2_{pq} B_{ij})} \exp{(-i \Delta m^2_{ij}}A_{pq})\Big)\Big],
\end{eqnarray}
where 
\begin{equation}
    |N|^2=\left(N_{path}+\sum_i |U_{\alpha i}|^2 \sum_{q>p}2 \cos{(\Delta b^2_{pq}B_{ii})}\right)^{-1}.
\end{equation}
Note that we only consider motion on the equatorial plane ($\theta=\pi/2$). Therefore, $N_{path}=2$, and the general form of the probability can be written as follows
\begin{strip}
\begin{eqnarray}
    {\mathcal{P}}^{lens}_{\alpha \beta}&=&|N|^2 \left[2 \sum_i |U_{\beta i}|^2|U_{\alpha i}|^2(1+\cos{(\Delta b^2 B_{ii}}))+\sum_{i,j\not=i}U_{\beta i}U^{*}_{\beta j}U_{\alpha j}U^{*}_{\alpha i}\big[\exp{(-i\Delta m^2_{ij} A_{11})}+\right. \nonumber \\
    &+&\exp{(-i\Delta m^2_{ij} A_{22})}+ 2 \left.\cos{(\Delta b^2 B_{ij})}\exp{(-i\Delta m^2_{ij}A_{12})} \big]\frac{}{}\right],
\end{eqnarray}
\end{strip}
where the normalization factor now is
\begin{equation}
     |N|^2=\left(2+2\sum_i |U_{\alpha i}|^2  \cos{(\Delta b^2_{pq}B_{ii})}\right)^{-1}.
\end{equation}
In this paper, a simple toy model of two neutrino flavors ($\nu_e\rightarrow\nu_{\mu}$) is considered to obtain a quantitative treatment of neutrino lensing in the spacetime of the Schwarzschild BH surrounded by a Dehnen-type DM halo. After that, the oscillation probability can be written as
\begin{eqnarray}\label{eq:prob}
    {\mathcal{P}}^{lens}_{e \mu} &=& |N|^2 \sin^2{2 \alpha}\Bigg[\sin^2{\left(\Delta m^2 \dfrac{A_{11}}{2}\right)}+\nonumber \\
&+& \sin^2{\left(\Delta m^2 \dfrac{A_{22}}{2}\right)}-\cos{(\Delta b^2 B_{12})}\cos{(\Delta m^2 A_{12})}+ \nonumber \\
   &+&\dfrac{1}{2}\cos{(\Delta b^2 B_{11})}+\dfrac{1}{2}\cos{(\Delta b^2 B_{22})}\Bigg],
\end{eqnarray}
and the normalization constant reduces to
\begin{equation}
        |N|^2 =\frac{1}{2}(1+\cos^2{\alpha}\cos{(\Delta b^2 B_{11})} 
        +\sin^2{\alpha}\cos{(\Delta b^2 B_{22})})^{-1}.
\end{equation}
where $\Delta b^2=\Delta b_{12}^2$ and $\Delta m^2=\Delta m_{21}^2$. 
Note that ${\cal P}_{e\mu}^{\text{lens}}$ remains unchanged under the interchange of the impact parameters $b_1$ and $b_2$. However, it depends separately on the quantities $B_{11}$ and $B_{22}$, which are defined from the Eq.~(\ref{Eq:AB}). The terms $B_{11}$ and $B_{22}$ are proportional to the absolute squared masses of the individual neutrino eigenstates. As a result, normal ($\Delta m^2>0$) and inverted ($\Delta m^2<0$) mass ordering, which correspond to different assignments of the nonzero mass to the eigenstates, lead to different oscillation probabilities. This feature does not appear in two-flavor neutrino oscillations in flat spacetime. Accordingly, it disappears in the absence of lensing ($\Delta b^2 = 0$) or for maximal mixing $\alpha = \pi/4$ (see,~\cite{Swami_2020}).

\subsection{Numerical results for two flavor toy model}

In this part, we investigate the effect of the DM halo parameters on the oscillation probability. For a better understanding, we demonstrate the schematic diagram of weak lensing of neutrinos in the spacetime of the Schwarzschild BH surrounded by a Dehnen-type DM halo in Fig.~\ref{fig:diagram}. Here, $r_S(x,y)$ and $r_D(x,y)$ refer to the distances from the massive object to the source and detector in a Cartesian plane $(x,y)$. Another $(x',y')$ coordinate system can be considered. This coordinate system is obtained by rotating the $(x,y)$ system at an angle $\varphi$. Therefore, we can write the following relations between these coordinate systems as follows.
\begin{align}
x'=x\cos{\varphi}+y\sin{\varphi}\,, \quad y'=-x\sin{\varphi}+y\cos{\varphi}\,.
\end{align}
After that, we can write the deflection angle in the following form
\begin{equation}
\delta\sim \frac{y'_D-b}{x'_D}=-\frac{2R_x}{b}\,,
\end{equation}
where 
\begin{equation}
    R_x=\frac{2M+16\pi\rho_a r_a^3}{1-32\pi \rho_a r_a^2}\,.
\end{equation}
If we consider $\sin{\varphi}=b/r_S$, we can obtain the following equation
\begin{equation}\label{Eq:impactp}
    (2 R_x x_D+b y_D)\sqrt{1-\dfrac{b^2}{r_S^2}}=b^2\left(\dfrac{x_D}{r_S}+1\right)-\dfrac{2 R_x b y_D}{r_S}.
\end{equation}
The solution to the above equation yields the impact parameter as a function of $r_S$, $R_x$, and $(x_D,y_D)$. Using the Sun-Earth system as an example, we model the Sun with a geometry of the Schwarzschild BH surrounded by a Dehnen-type DM halo and place a detector at Earth's location ($r_D = 10^8 km$). The model assumes a source of relativistic neutrinos, with a characteristic energy $E_0 = 10 \ MeV$, positioned behind the Sun at a distance $r_S = 10^5r_D$. As the detector moves in its circular orbit, we calculate two real solutions of Eq.~(\ref{Eq:impactp}) (with impact parameters $b_1$ and $b_2$) for every $\varphi$ by solving the key equation numerically. In this numerical analysis, the probability of neutrino oscillations is evaluated only for those values of $b_p$ that satisfy the condition $R_x \ll b_p \ll r_D$. That is, neutrinos are assumed to travel at sufficiently large distances from the matter source so that the weak lensing approximation holds, while the detector is positioned much farther away compared to the impact parameter. The other parameters are $M=M_{\odot}$ and $|\Delta m^2|=10^{-3} \ eV^2$. It should be noted that the values used here are for conceptual demonstration; any practical application requires precise numerical parameters for the model's geometry. Finally, we plot the dependence of the oscillation probability on the azimuthal angle $\varphi$ for different values of the spacetime parameters by considering the normal and inverted hierarchies in Figs.~\ref{fig:prob1},~\ref{fig:prob2} and~\ref{fig:prob3}. Here, the corresponding parameters are $M=1M_{\odot}$, $\Delta m^2=10^{-3} \ eV^2$, and the mixing angle is $\alpha=\pi/6$.  It can be observed from these figures that the values of the oscillation probability slightly shift toward the higher values of the azimuthal angle $\varphi$ with increasing spacetime parameters. In addition, there is a slight increase/decrease under the influence of the spacetime parameters when we are considering the inverted/normal hierarchy.   

\begin{figure*}[!htb]
\centering
\includegraphics[scale=0.5]{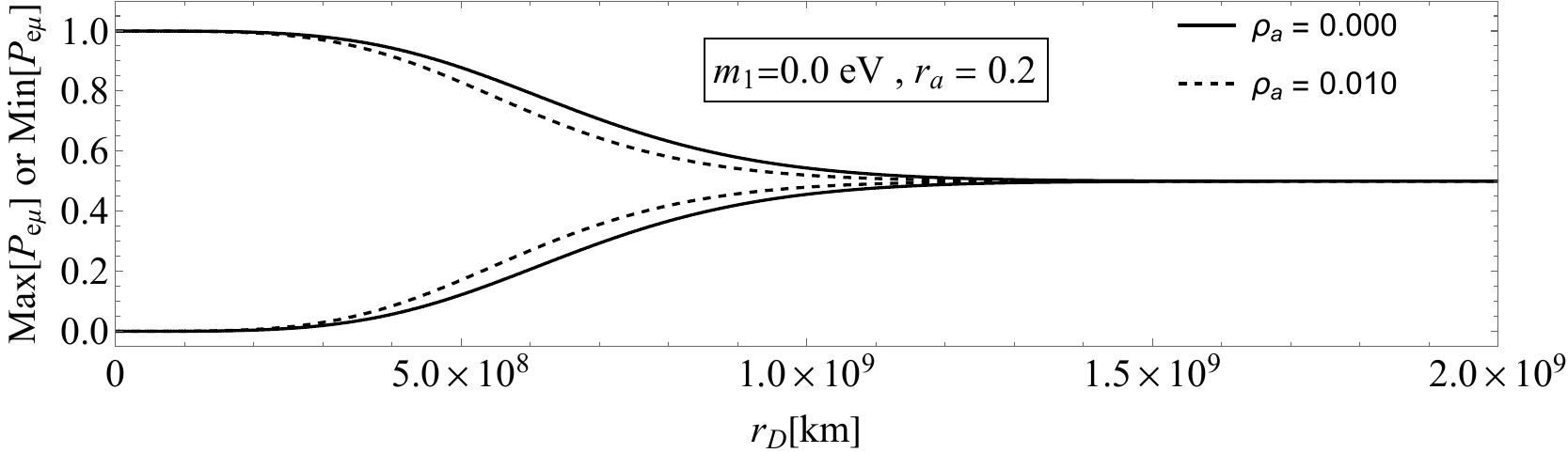}
\includegraphics[scale=0.5]{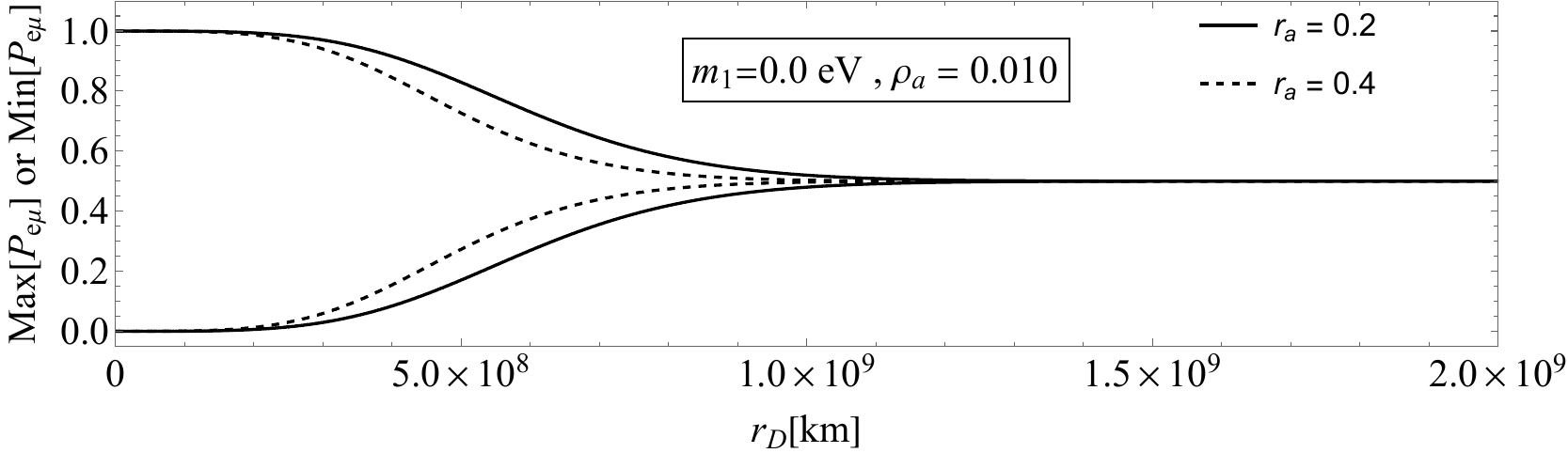}
\caption{The plot shows the dependence of the maximum and minimum transition probability envelop on the $r_D$ for the different values of the $\rho_a$ (top panel) and $r_a$ (bottom panel). The other parameters are the same as given in the caption of Fig.~\ref{fig:damping}. Here, the mixing angle is equal to $\alpha=\pi/4$.}
\label{fig:transition}
\end{figure*}
\section{Decoherence}\label{sec:6}

In this part, we investigate the effect of decoherence on the neutrino oscillations. Here, we assume that neutrinos are characterized by Gaussian wave packets instead of the plane-wave approximation. In a gravitational field, the decoherence length depends on the proper time along the particle's path from source to detector. Under weak-field conditions, the proper time interval for a given proper distance is shorter than its flat spacetime equivalent. Thus, in the presence of a compact object, neutrino wave packets need to cover a longer spatial distance to accumulate the same proper time before decoherence occurs. The probability can be written in terms of the wave packet as follows (see~\cite{Swami_2021})
\begin{equation}\label{prob}
    \mathcal{P}_{\alpha\beta} = \frac{\sum_{i,j} U_{\beta i}^* U_{\alpha i} U_{\beta j}^* U_{\alpha j} \sum_{p,q} e^{-i \Phi_{i j}^{p q}}   e^{- X_{i j}^{p q}}}{\sum_{i} U_{\alpha i} U_{\alpha i}^* \sum_{p,q}e^{-i \Phi_{ii}^{p q}} \ e^{- X_{ii}^{p q}}} ,
\end{equation}
with 
\begin{align}\label{dPhase1}
    \Phi_{ij}^{pq}= \left(\Phi_{i}^p-\Phi_{j}^q\right)- \frac{\Bar{\sigma}^2}{\sigma_D^2 }\left(\Vec{p}^{\,D}-\Vec{p}^{\,S}\right) \left(\Vec{\textbf{X}}_i^p-\Vec{\textbf{X}}_j^q\right) \ ,
\end{align}
and
\begin{align}\label{dPhase2}
     \textbf{X}_{ij}^{pq}= \frac{1}{2} \Bar{\sigma}^2 \left(|\Vec{\textbf{X}}_i^p|^2+|\Vec{\textbf{X}}_j^q|^2\right). 
\end{align}
where $\Bar{\sigma}^2=\sigma_D^2 \sigma_S^2/\left(\sigma_D^2+\sigma_S^2\right)$ and $\Vec{\textbf{X}}_i^p = \partial_{\Vec{p}}\Phi_{i}^p $. Using the expression for phase $\Phi_i^p$, we can easily find $\Vec{X}_i^p$ as follows
\begin{eqnarray}
       |\Vec{X}_i^p|^2 &\simeq& -\dfrac{m_i^4}{4 E_{loc}^4  {\cal A}(r_S)}(r_S+r_D)^2\Bigg(1-\frac{b_p^2}{2 r_Sr_D}(1-32 \pi  \rho _a r_a^2)+\nonumber \\
&+&\frac{2M}{r_S+r_D}\frac{(1+8 \pi  \rho_a r_a^3)}{1-32 \pi  \rho_a r_a^2}\Bigg)^2 \simeq \nonumber \\
       & \simeq& -\dfrac{m_i^4}{4 E_{loc}^4  {\cal A}(r_S)}(r_S+r_D)^2\Bigg(1-\frac{b_p^2}{2 r_Sr_D}(1-32 \pi  \rho _a r_a^2)+\nonumber \\
&+&\frac{4M}{r_S+r_D}\frac{(1+8 \pi  \rho_a r_a^3)}{1-32 \pi  \rho_a r_a^2}\Bigg)
\end{eqnarray}
at the leading order in ${m_i}/{E_{loc}}$, where $E_{loc}$ refers to the energy observed by a local observer. It can be written as~\cite{Swami_2021}
\begin{equation}
    E_{loc}=E_{loc}(r_S)=\frac{E_0}{\sqrt{-{\cal A}(r_S)}}.
\end{equation} 
The decoherence can be measured by using the effective damping factor $D_{ij}^{pq}=\textbf{X}_{ij}^{pq}-\textbf{X}_{\hat{i}\hat{i}}^{\hat{p}\hat{q}}$, and considering that the gravitating object is located between the source and the detector. In this scenario, neutrinos may follow two distinct classical trajectories, labelled $p=1$ and $q=2$, which are characterized by their respective impact parameters $b_1$ and $b_2$. If we align the x-axis with the line that joins the neutrino source and the matter source, we can choose impact parameters that satisfy $b_1\leq b_2$ for $y\geq0$, and order the neutrino masses in ascending sequence as $m_1<m_2<...<m_n$. After that, we can obtain the damping factor for $y\geq0$ in the following form
\begin{eqnarray}\label{Dij}
    D_{ij}^{pq}&=&\textbf{X}_{ij}^{pq}-\textbf{X}_{11}^{11} \simeq \nonumber \\
    &\simeq&-\frac{\Bar{\sigma}^2 \left( r_S+r_D \right)^2 }{8 E_{loc}^4  {\cal A}(r_s)} \Bigg(1+\dfrac{4 M}{r_S+r_D}\frac{(1+8 \pi  \rho_a r_a^3)}{1-32 \pi  \rho_a r_a^2}\Bigg) \times \nonumber \\
    &\times&\Bigg[m_i^4\Bigg(1-\frac{b_p^2}{r_S r_D}(1-32 \pi  \rho _a r_a^2) \Bigg)+ \nonumber \\
    &+& m_j^4\Bigg(1-\frac{b_q^2}{r_S r_D}(1-32 \pi  \rho _a r_a^2) \Bigg)-\nonumber \\
    &-& 2m_1^4\Bigg(1-\frac{b_1^2}{r_S r_D}(1-32 \pi  \rho _a r_a^2) \Bigg)\Bigg]\ .
\end{eqnarray}
Decoherence can arise in two ways: (a) from the mass difference between the lightest and the second-lightest neutrino mass eigenstates and (b) from a path difference, even in the case $i=j=1$. The latter is negligible because it arises only in sub-leading order in ${m_{i,j}}/{E_{loc}}$. It is important to note that the geometry of the Schwarzschild BH surrounded by a Dehnen-type DM halo can be distinguished from that of Schwarzschild through the measurement of the damping factor or the decoherence length. It is possible to estimate the decoherence length for the Sun-Earth system~\cite{Swami_2021}. The parameters are considered as follows: $R_x=3 \ km$, $E_{loc}=10 \ MeV$, and $r_S=10^5 r_D$. In addition, the co-linear case, in which the source, detector, and the massive object lie on the same line, is considered for simplicity. To provide more information, we visualize the results obtained in Fig.~\ref{fig:damping}. The left panel illustrates the $D^{11}_{12}$ dumping factor as a function of $r_D$ for $m_1=0 \ eV$ and $m_1=0.1 \ eV$. Note that the left panel corresponds to the Schwarzschild BH case. Moreover, the effect of the DM halo's parameters on the damping factor is demonstrated by the middle and right panels. From these panels, it can be observed that the damping factor values for fixed values of $r_D$ increase with the increase of the DM halo parameters. Furthermore, we plot the dependence of the maximum and minimum transition probability envelope on the $r_D$ for the different values of the spacetime parameters in Fig.~\ref{fig:transition}. It should be noted that we consider the range of $r_D$ from $10^8 \ km$ to $2\cdot10^9 \ km$ with the interval $\Delta r_D=1.9\cdot10^6 \ km$. Therefore, the plot has $2000$ data points. The plot shows that after a certain distance, neutrino coherence is lost. At this point, the influence of the DM halo parameters becomes negligible compared to the dominant effect of the absolute neutrino masses. It is worth noting that the probability saturates when the neutrino travels a large enough distance, and the probability approaches the value determined by the mixing matrix element, i.e., $P_{\alpha \beta}\rightarrow |U_{\beta  1}|^2$ (see,~\cite{Swami_2021}).  In the considered case, it is equal to $P_{e\mu} \rightarrow \sin^2{\alpha}=\frac{1}{2}$. It can also be seen from the Figs.~\ref{fig:damping} and~\ref{fig:transition} probability saturation occurs at radial distance $r_D$ where $D^{11}_{22}\geq2$. Once $r_D$ exceeds the decoherence length, interference effects from lensing of neutrinos diminish, and the probability saturates to the above value in this case.

\section{Conclusions}\label{sec:summary}

In this paper, we investigate the neutrino flavor oscillation in the spacetime of the Schwarzschild BH surrounded by a Dehnen-type DM halo. Firstly, we briefly review the Schwarzschild BH surrounded by a Dehnen-type DM halo. In Fig.~\ref{fig:region}, we plot the phase diagram indicating the existence of the BH in two different slices of the parameter space. We then study the effect of the DM halo parameters by addressing the event horizon radius. Fig.~\ref{fig:horizon} shows the dependence of the radius of the event horizon on the density of the DM halo $\rho_a$ for different values of $r_a$. It can be observed from this figure that the values of the event horizon radius increase with increasing density and radius of the DM halo. After that, we study the propagation of neutrinos in the gravitational weak field of the Schwarzschild BH surrounded by a Dehnen-type DM halo by considering two types of propagation, which are radial and non-radial. We obtain the analytical expressions for the phase and probability of the neutrino oscillations. We consider a simple toy model of two neutrino flavors ($\nu_e\rightarrow\nu_{\mu}$) to visualize our results. It is worth noting that we explore two cases, which are normal ($\Delta m^2>0$) and inverted ($\Delta m^2<0$) hierarchies. Our numerical results for the probability of neutrino oscillations are demonstrated in Figs.~\ref{fig:prob1},~\ref{fig:prob2}, and~\ref{fig:prob3}. Moreover, we obtain the analytic expression for the damping factor by using the equation for the phase of neutrino oscillations. Fig.~\ref{fig:damping} shows the damping factor as a function of $r_D$, which is the distance between the massive object and the detector for different values of the spacetime parameters. Our results demonstrate that the damping factor values for fixed $r_D$ increase with increasing spacetime parameters. Finally, we plot the maximum and minimum transition probability envelope as a function of $r_D$ for different values of the spacetime parameters in Fig.~\ref{fig:transition}. It can be observed from this figure that neutrino coherence is lost after a certain distance. 

Now, we have a chance to answer the question of this paper, which is \textit{Can a Dehnen-type dark matter halo affect the neutrino flavor oscillations?}. While one might qualitatively expect an effect due to the additional gravitational mass from the DM halo, the purpose of this study is to quantify and distinguish how the resulting geometry deviates from that of a standard Schwarzschild BH, and how these differences manifest in neutrino oscillations. The performed analysis indicates that, in principle, sensitive neutrino detectors could differentiate between the Schwarzschild BH with a Dehnen-type halo and one without it.  However, currently, carrying out such experiments remains unattainable. Nevertheless, the model, while simplified, demonstrates that in principle one could extract important information about the geometry from neutrino observations.

\section*{Acknowledgement}

MA warmly thanks Daniele Malafarina and Hrishikesh Chakrabarty for their valuable comments and discussions on this topic. We warmly thank the anonymous referee for valuable comments and discussions that helped to improve the accuracy and quality of the presentation of the manuscript. This research was funded by the National Natural Science Foundation of China (NSFC) under Grant No. U2541210.

\section*{Data Availability Statement}

This manuscript has no associated data.

\section*{Code Availability Statement}
This manuscript has no associated
code/software.

\bibliographystyle{spphys}
\bibliography{ref}

\end{document}